\documentclass[pra,superscriptaddress,floatfix,amsmath,nofootinbib,twocolumn,amssymb]{revtex4-1}

\usepackage[abs]{overpic}
\usepackage{bm}
\usepackage{verbatim}
\usepackage{mathrsfs}
\usepackage{color}
\usepackage{paralist}
\usepackage{hyperref}

\begin{document}

\title{Dynamical Casimir effect in Circuit QED for Nonuniform Trajectories}
\author{Paulina Corona-Ugalde}
\email{pcoronau@uwaterloo.ca}
\affiliation{Institute for Quantum Computing, University of Waterloo, Waterloo, Ontario, N2L 3G1, Canada}
\affiliation{Department of Physics \& Astronomy, University of Waterloo,  Ontario Canada N2L 3G1}
\author{Eduardo Mart\'in-Mart\'inez}
\email{emmfis@waterloo.ca}
\affiliation{Institute for Quantum Computing, University of Waterloo, Waterloo, Ontario, N2L 3G1, Canada}
\affiliation{Department of Applied Mathematics, University of Waterloo, Waterloo, Ontario, N2L 3G1, Canada}
\affiliation{Perimeter Institute for Theoretical Physics, Waterloo, Ontario, N2L 2Y5, Canada}
\author{C.M. Wilson}
\email{chris.wilson@uwaterloo.ca}
\affiliation{Institute for Quantum Computing, University of Waterloo, Waterloo, Ontario, N2L 3G1, Canada}
\affiliation{Department of Physics \& Astronomy, University of Waterloo,  Ontario Canada N2L 3G1}
\author{Robert B. Mann}
\email{rbmann@uwaterloo.ca}
\affiliation{Institute for Quantum Computing, University of Waterloo, Waterloo, Ontario, N2L 3G1, Canada}
\affiliation{Department of Physics \& Astronomy, University of Waterloo,  Ontario Canada N2L 3G1}

\begin{abstract}
We propose a generalization of the superconducting circuit simulation of the dynamical Casimir effect where we consider relativistically moving boundary conditions following different trajectories. We study the feasibility of the setup used in the past to simulate the dynamical Casimir effect to reproduce richer relativistic trajectories differing from purely sinusoidal ones. We show how  different relativistic oscillatory trajectories of the boundaries of the same period and similar shape produce a rather different spectrum of particles characteristic of their respective motions.
\end{abstract}

\maketitle

\section{Introduction}

The observation of the Unruh effect, either directly or in analogue systems, is one of the experimental cornerstones of quantum field theory in curved spacetimes and relativistic quantum information.  Since field quantization schemes associated with inertial and accelerated observers are not equivalent \cite{Birrell1984},  observers uniformly accelerating in what inertial observers regard as a  vacuum will detect a thermal bath of particles \cite{Unruh1976}.   The temperature $T$ of this thermal bath is predicted to be proportional to the magnitude ${a}$ of the proper acceleration of the detector.   The breadth of context in which this effect  has been predicted and derived -- including axiomatic quantum field theory \cite{Sewell1982}, via Bogoliubov transformations \cite{Birrell1984}, and the response rates of noninertial particle detectors both perturbatively \cite{Birrell1984} and nonperturbatively \cite{Brown2012,Bruschi2012,Hu:2012jr,Doukas:2013noa} -- have led physicists to regard it as a universal phenomenon.

Yet generalizations of this effect to other (nonequilibrium) regimes, such as nonuniformly accelerated trajectories \cite{Mann:2009dma,Ostapchuk:2011ud}  and short times  are not completely understood, even  from a theoretical point of view \cite{Doukas:2013noa,Wilson,Brown2012,Brenna2013,Brenna:2015fga}.  Recently it has been shown \cite{Brown2012}  that within optical cavities in (1+1)-dimensions an accelerated detector equilibrates to a thermal state whose temperature is proportional to its acceleration. 
Provided the detector is allowed enough interaction time, this effect holds independently of the cavity boundary conditions \cite{Brenna2013}, though for sufficiently short timescales (still long enough to satisfy the KMS condition) the temperature  decreases with acceleration in certain parameter regimes \cite{Brenna:2015fga}.

  In the classic Unruh Effect a uniformly accelerated detector in an inertial vacuum measures thermal radiation at the Unruh temperature $T_U$
\begin{equation}
\label{Unruh}
k_B T_U= \frac{\hbar a}{2\pi c}
\end{equation} 
Amongst the  problems one encounters when trying to experimentally detect this effect, the two main ones are (a) an inability to eternally accelerate anything (hence uniformity of acceleration cannot always hold) and (b) in practical terms,  difficulty  in accelerating a physical detector, such as a 2-level atom, with sufficient control. For these reasons, it would be extremely useful to have a quantum simulation of these phenomena.  However its implementation  requires some care.

The first problem involves overcoming the idealization of uniformity of acceleration by considering generalizations to nonuniformly accelerating trajectories. Under general conditions, a particle detector undergoing a general nonintertial trajectory will register a coloured noise that turns thermal only under the limiting conditions of uniform acceleration. The natural setting to consider is  oscillatory motion, which is more convenient for experimental implementations and extremely interesting from a theoretical point of view. A recent analysis of detectors undergoing various kinds of oscillatory motion \cite{Doukas:2013noa} found that in general such
detectors responded to the vacuum fluctuations of a quantum field and experienced a constant effective temperature at late times in these out of equilibrium conditions.  Three kinds of oscillatory motion --  sinusoidal motion, sinusoidal acceleration  and alternating uniform acceleration (AUA) -- were considered, and the effective temperature for each was found to depend more strongly on the geometry of the worldline  than on the instantaneous proper acceleration. The behaviour of  their steady state temperature was seen to be more similar to each other than to that of the Unruh temperature of an idealized uniformly accelerated detector provided
the time scale of the detector's response was longer than the period of the oscillatory motion.
 
The second problem, that of the difficulty of (relativistically) accelerating a detector (even under the restriction to oscillatory motion) 
can be addressed by considering an inertial detector and a moving reflective boundary (or mirror).  For a mirror that uniformly accelerates at late times, the detector experiences the  same thermal radiation as predicted in the original Unruh effect. 

Here we analyze the behaviour of oscillatory boundaries in the context of circuit-QED. Superconducting circuits offer an ideal testbed for implementation of this kind of experiment. This idea is reinforced by the fact that the Unruh effect is strongly connected with the dynamical Casimir effect. As such, it may be possible to modify  settings  where the latter is simulated \cite{Wilson,DCE,DCE2} to study  particle creation due to effectively relativistic noninertial trajectories of boundary conditions.  We consider the same three motions as in ref. \cite{Doukas:2013noa}, and compute the number of photon quanta emitted as a result.
 
In section \ref{RTMB} we review the different relativistic moving boundary conditions that we investigate, and discuss in \ref{sect-cQED} how to interpret these in the context of circuit QED. We compute in section  \ref{NPRT} the output number of photons for each trajectory for a variety of realistic values of the input parameters, and discuss this differs for different relativistic trajectories in section \ref{results}.  We conclude with a brief discussion of the implications of our work for detecting the Unruh effect.
 
\section{Relativistic Trajectories and Moving Boundaries}\label{RTMB}

Our aim is to simulate the different relativistic motions of a boundary using a slight modification of the setup previously used to simulate the dynamical Casimir effect \cite{Wilson,DCE,DCE2}, to compute the photon emission spectrum, and to extract from this an effective temperature.   This setup -- a Coplanar Waveguide (CPW) terminated by a SQUID -- will be used to simulate a moving boundary described by $x=z(t)$, where $t$ is the coordinate time in the lab frame. The relativistic trajectories that we will simulate are the ones studied in \cite{Doukas:2013noa}: sinusoidal motion (SM), sinusoidal acceleration (SA), and alternating uniform acceleration (AUA). 
The periodic boundary motion is in one dimension, and we define its ``directional proper acceleration"
as its  (positive definite) proper acceleration multiplied by the sign of  the spatial component of the 4-acceleration. 
We denote the periodicity of the motion as $\omega_d$, anticipating that the external driving  flux that we will use to simulate these trajectories will also have the same natural frequency, as we will see  in detail later on.

Sinusoidal motion is one for which the 4-position of the boundary is given by
\begin{equation}\label{Sinusoidal}
z_{\textsc{SM}}^\mu (t)=(t,0,0,-R \cos (\omega_d t))
\end{equation}
where   $R$ is the oscillation amplitude, and $\omega_d$ is the oscillation frequency in coordinate time. In order for the motion to remain subluminal we must have $R\omega_d< 1$ in units with $c=1$. The proper time $\tau$ of the detector is $\tau=\omega^{-1} E(\omega t, (R\omega)^2)$, where $E(\phi, m)$ is an elliptic integral of the second kind. The directional proper acceleration is 
$$
\alpha^{}_{\textsc{SM}}(t) =  R\omega_d^2 \frac{\cos{\omega_d t}}{(1-(\frac{R\omega_d}{v})^2 \sin^2{\omega_d t})^{3/2}}
$$
whereas the proper acceleration $a^{}_{\rm SM}(t)= |\alpha^{}_{\rm SM}(t)|$.  Note that for $R\omega_d \ll 1$ the acceleration is proportional to the position, as expected for nonrelativistic motion.  The oscillation period is
 $t_p=2\pi/\omega_d$ (or  $\tau_p= \omega_d^{-1} E(2\pi, (\frac{R\omega_d}{{v}})^2)$ in proper-time). The time-averaged proper acceleration (over one oscillation period) is:
\begin{eqnarray}
  \bar{a} &=& {v\omega_d \tanh^{-1}\frac{R \omega_d}{v} \over E\left(\frac{R^2\omega_d^2}{v^2}\right)}.
\end{eqnarray}
The acceleration profile for the SM worldline develops extra peaks due to relativistic dilation effects that create periodic positive double-peaks (or 'kicks') in the two-point correlation function for the photon field \cite{Doukas:2013noa}.

Sinusoidal acceleration (SA) (employed in a experimental proposal by Chen and Tajima
\cite{Chen1999a}, in which a particle of mass $m$ and charge $e$ is placed at one of the magnetic nodes of an EM standing wave with frequency $\omega_d$ and ampitud $E_0$)
is described by the worldline
\begin{equation}\label{Chen}
z_{\textsc{SA}}^\mu (t)=\left[t,0,0,-\frac{v}{\omega_d} \arcsin\left(\frac{2\frac{ \alpha}{v \omega_d} \cos(\omega_d t)}{\sqrt{1+4\left(\frac{\alpha}{v\omega_d}\right)^2}}\right)\right]
\end{equation}
with directional proper acceleration 
$$
\alpha^{}_{\textsc{SA}}(t) = 2 \alpha  \cos \omega_d t
$$
where $\alpha=\frac{e E_0}{m}$ has units of acceleration and the proper time of the detector $\tau$ is related to the coordinate time $t$ by $ \tau(t)  = \omega_d^{-1}F(\omega_d t, -4\alpha^2/v^2\omega_d^2)$, where $F(\phi, m)$ is the elliptic integral of the first kind.  The oscillation period of this worldline  is $t_p= 2\pi/\omega_d$ or a proper time period of $\tau_p= \omega_d^{-1}F(2 \pi, -4\alpha^2/v^2\omega_d^2)$. The time-averaged proper acceleration reads
\begin{equation}
  \bar{a} = {v \omega_d \sinh^{-1}\left(2\frac{\alpha}{v\omega_d}\right) \over F\left( \pi/2, -4\left(\frac{\alpha}{v\omega_d}\right)^2\right)}.
\end{equation}
and for low accelerations ($|\alpha| \ll v\omega_d$)
and nonrelativistic velocities we obtain $z_{\rm SA}\sim z_{\rm SM}$.

For Alternating Uniform Acceleration (AUA) the trajectory of the boundary (parametrized in the accelerated observer's proper time) is
\begin{widetext}
\begin{equation}\label{AUA}
z_{\textsc{aua}}^\mu (\tau)=\left[\frac{v^2}{a}\left[\sinh \frac{a}{v}\left(\tau -\frac{n \tau_p}{2}\right)+2n\sinh \frac{a \tau_p}{4 v}\right],0,0,\frac{(-1)^n v^2}{a}\left[\cosh \frac{a}{v}\left(\tau -\frac{n \tau_p}{2}\right)+\lbrace(-1)^n-1\rbrace\cosh \frac{a \tau_p}{4 v}\right]\right]
\end{equation}
and so it experiences constant acceleration $a$ that periodically alternates in sign
\begin{equation}
  a^\mu_{\textsc{AUA}}(\tau)= \left(a \sinh \frac{a}{{v}}\left[\tau- {n\tau_p\over 2}\right], 0, 0, (-1)^{n}a \cosh \frac{a}{{v}}\left[\tau- {n\tau_p\over 2}\right]\right).
\end{equation}
where  $n(\tau)\equiv {\textrm{floor}}\left( {2\tau\over\tau_p} +{1\over 2}\right)$,  with ${\textrm{floor}}(x)$
the largest integer less than or equal to $x$. We consequently have $\bar{a}=a$.
 \end{widetext}
 
\begin{figure}[htp]
    \includegraphics[scale=.6]{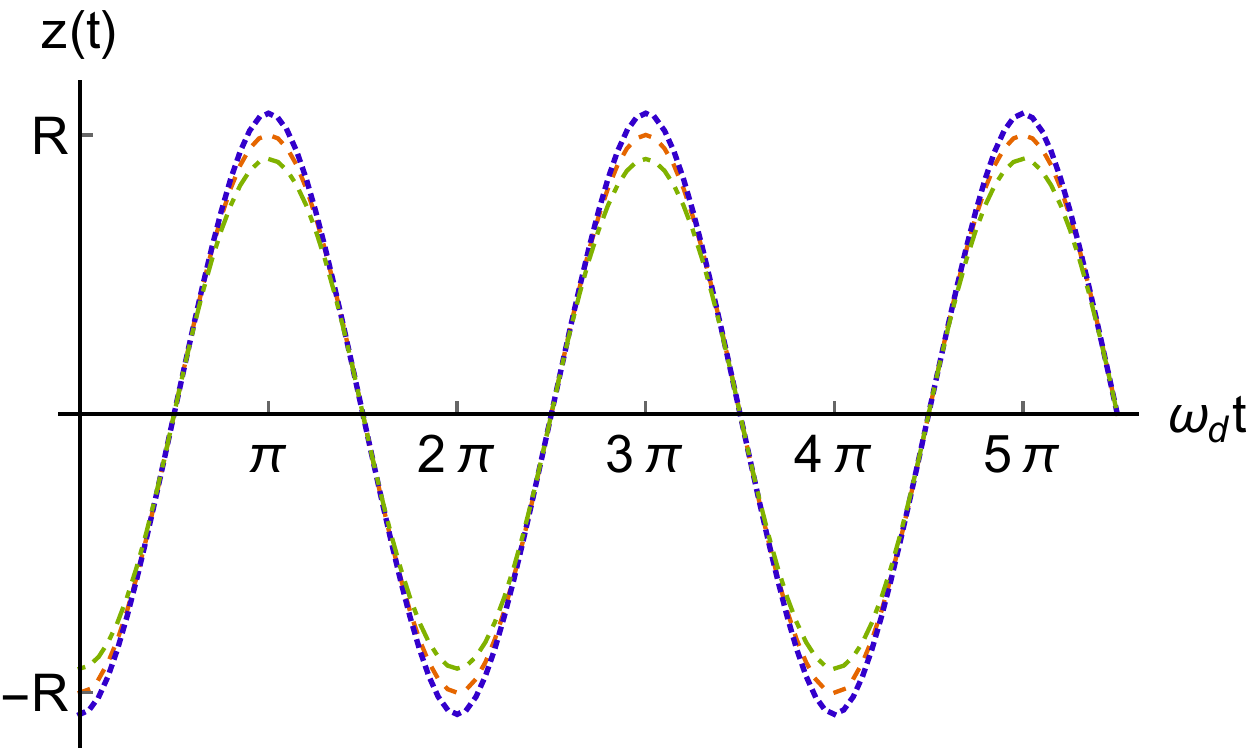}
     \includegraphics[scale=.6]{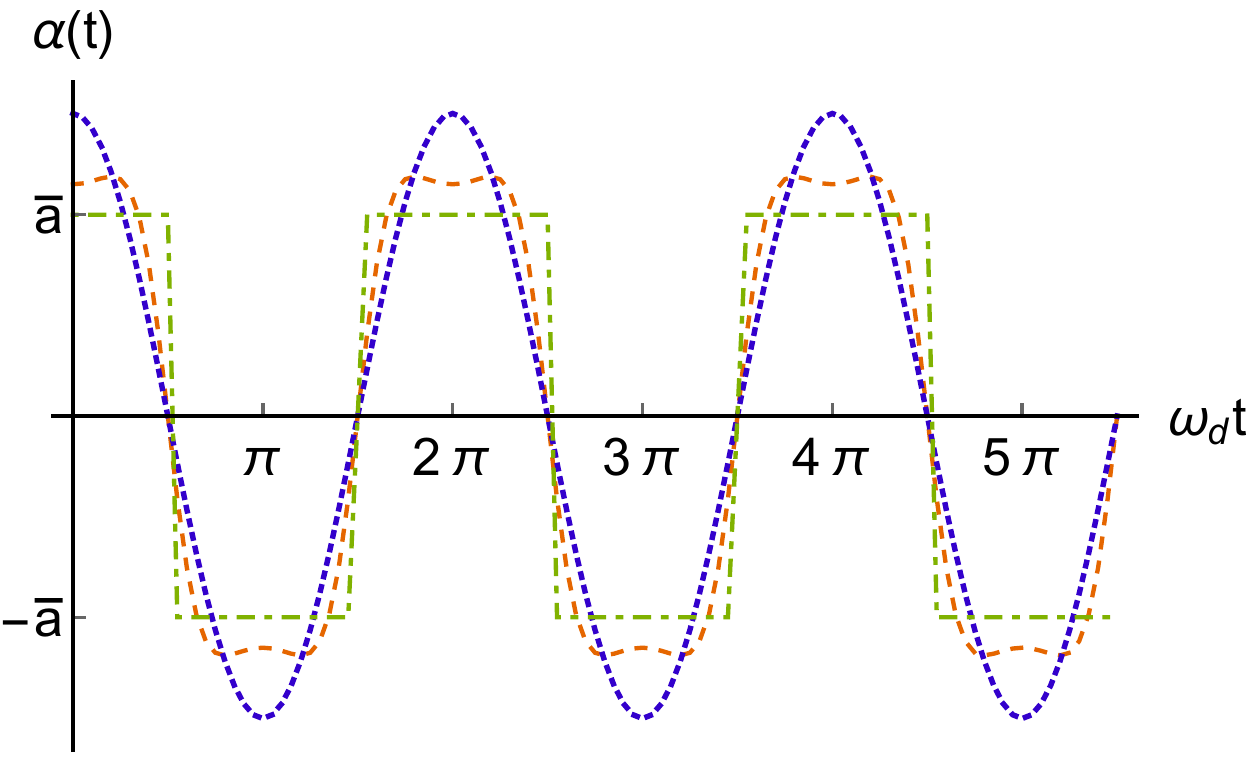}
    \caption{(Upper) Position as a function of time. (Lower) Directional acceleration as a function of time. In both cases the trajectories are distinguished as follows: Sinusoidal motion (Red, dashed), Sinusoidal Acceleration(Blue, dotted), Alternating Uniform Acceleration(Green, dot-dashed). The average acceleration for all trajectories is $\bar{a}=1.2\times 10^{19}\text{m}/\text{s}^{-2}$ and the driving frequency is $\omega_d/2\pi=28$ GHz.}
    \label{doukas}
    \end{figure}

We illustrate in Figure \ref{doukas}  the  position (upper) and  proper acceleration (lower) as functions of time.

\section{cQED setup}
\label{sect-cQED}

To simulate these boundary motions we make use of the setup \cite{DCE2} (illustrated in Figure $2$ of \cite{DCE2}), where a Superconducting Quantum Interference Device (SQUID) modulates the boundary condition of a Coplanar Waveguide (CPW). The CPW is at $x < 0$ and the SQUID is at $x=0$.

The dynamical flux away from the SQUID is found to respect the Klein-Gordon equation
\begin{equation}\label{eqmotion}
\frac{\partial^2}{\partial t^2}\Phi(x,t)-v^2\frac{\partial^2}{\partial x^2}\Phi(x,t)=0
\end{equation}
with propagating velocity $v=1/ \sqrt{C_0 L_0}$, where $C_0$ and $L_0$ are the capacitance and inductance per unit length of the CPW.

The equation of motion at the boundary is
\begin{eqnarray}
0 &=& C_J\ddot{\Phi}(0,t)+\left(\frac{2\pi}{\phi_0}\right)^2E_J(t)\Phi(0,t)+\frac{1}{L_0}\frac{\partial \Phi(x,t)}{\partial x}\bigg\vert_{x=0}\nonumber\\
&\simeq& \Phi(0,t)+\frac{1}{L_0 E_J(t)}\left(\frac{\phi_0}{2\pi}\right)^2\frac{\partial \Phi(x,t)}{\partial x}\bigg\vert_{x=0}
\label{boundary}
\end{eqnarray}
where $\phi_0=\frac{h}{2e}$ is the magnetic flux quantum, $C_J$ is the capacitance of the symmetric SQUID, which has a small enough loop (so that self-inductance is neglected) and operates in the phase regime, and $E_J(t)=E_J(\Phi_{\text{ext}}(t))$ is the tunable Josephson energy whose arbitrary time dependence can be given by controlling $\Phi_{\text{ext}}$, the external flux threading through the SQUID. The second equality follows under the assumption that the SQUID plasma frequency is much larger than any other frequencies in the circuit. This boundary can be tuned by the externally applied magnetic flux.

Remember that for a field terminated by a moving mirror, we would have the boundary condition
\begin{equation}\label{bcmotion}
\phi(t,Z(t)) = 0
\end{equation}
where $Z(t)$ is some prescribed trajectory.  

Note that the above equation can be written in the approximate form 
\begin{equation}\label{abcmotion}
\phi(t,x)\bigg\vert_{x=0} + (Z_0 -  z(t)) \frac{\partial\phi(t,x)}{\partial x }\bigg\vert_{x=Z_0} = 0
\end{equation}
upon expanding equation (\ref{bcmotion}) about the origin, where $Z(t) = Z_0 -  z(t)$ with $  z(t) \ll Z_0$.  Equation (\ref{boundary}) plays the role of the boundary condition  on the flux field in the CPW, and is designed to simulate the boundary condition (\ref{abcmotion}).  We remark  that the boundary condition \eqref{boundary} does not exactly correspond to a Dirichlet condition, and instead it is similar to it only in an approximate way.   In principle  identifying this simulation with the original perfect-mirror Dirichlet boundary condition employed in the classic literature on the dynamical Casimir effect \cite{Moore1,Fulling1,Dodonov1} can be problematic.  This is  because the condition \eqref{boundary} well approximates a pure Dirichlet condition at a moving boundary only when $|dz(t)/dt| \ll c$ \cite{inprep}. 
 
 In our case  this is not a concern  for two reasons. First, the dynamical Casimir effect does not require   strict use of a Dirichlet condition; indeed it occurs   for a general set of time dependent boundary conditions near relativistic regimes  \cite{generalizedDynamical}.
  Second,
 the boundary condition \eqref{abcmotion} (which is faithfully approximated by \eqref{boundary} for field frequencies much smaller than the SQUID plasma frequency \cite{JormJas}) produces the same particle spectrum (at leading order) as the pure Dirichlet condition.
Writing $V=\omega t+k_{\omega} x$ and $U=\omega t-k_{\omega} x$,
the full solution to  (\ref{eqmotion}) that respects  (\ref{bcmotion}) is given by
\begin{eqnarray}
&&\phi(t,x) = f(V) - f(p_+(U)) + g(U) - g(p_-(V)) \nonumber\\
&& \textrm{where} \quad  
\omega t\pm k_{\omega} z_\pm(t)  = p_\pm(\omega t\mp k_{\omega}z_\pm(t))
\label{solbc}
\end{eqnarray} 
in the case of two boundaries with trajectories   $V=p_+(U)$ and $U=p_-(V)$,
respectively determined in terms of the prescribed boundary motions $x=z_\pm(t)$.
 We shall set $f(V)=0$ as there is neither a left boundary nor  incoming
right-propagating signals, and write $z_-(t)=z(t)$.
 
  The general method for interpreting this equation 
(for left-moving modes that are reflected from the boundary) is to write (\ref{solbc}) as
\begin{eqnarray}\label{phi}
&&\Phi(x,t)=\sqrt{\frac{\hbar Z_0}{4 \pi}}\int_0^\infty \frac{d \omega}{\sqrt{\omega}} \left(a_{\text{in}}(\omega)e^{-i(-k_{\omega} x+\omega t)} \right.  \nonumber\\
&& \qquad\qquad \qquad   +a_{\text{out}}(\omega)e^{-i(k_{\omega} x+\omega t)}+\text{H.c} \Big)
\end{eqnarray}
where $Z_0=\sqrt{L_0/C_0}$ is the characteristic impedance and $k_\omega=|\omega|/v$ is the wave vector. The subscripts {\it out} and {\it in} in the operators stand for the direction in which the signals are propagating, with   $a_{\text{in}}(\omega) = \int dU  g(U) e^{-i {\omega}U}$ and we  interpret $a_{\text{out}}(\omega) = \int dV  g(p_+(V))  e^{i {\omega}V}$.  Rather than directly computing this latter integral,
we shall obtain $a_{\text{out}}(\omega)$ by
requiring the field (\ref{phi}) to satisfy the boundary condition \eqref{boundary}. 
After a Fourier transformation this   yields 
\begin{widetext}
\begin{equation}\label{resolver}
0=\left(\frac{2\pi}{\Phi_0}\right)^2\int_{-\infty}^\infty d\omega g(\omega,\omega ')[\Theta(\omega)(a_\omega^{\text{in}}+a_\omega^{\text{out}})
+\Theta(-\omega)(a_{-\omega}^{\text{in}}+a_{-\omega}^{\text{out}})^\dagger]-\omega '^2 C_J(a_{\omega '}^{\text{in}}+a_{\omega '}^{\text{out}})
+i \frac{k_{\omega '}}{L_0}(a_{\omega '}^{\text{in}}-a_{\omega '}^{\text{out}})
\end{equation}
\end{widetext}
where
\begin{equation}\label{solve}
g(\omega,\omega ')=\frac{1}{2 \pi}\sqrt{\frac{|\omega '|}{|\omega|}} \int_{-\infty}^{\infty} \!\!\!\text{d}t\, E_J(t) e^{-i(\omega-\omega ')t}
\end{equation}

 Consider an arbitrary driving motion $E_J(t)$ with  Fourier  decomposition
\begin{equation}\label{FouGen}
E_J(t)=\frac{a_0}{2}+\sum_n a_n \cos (\omega_d nt)+\sum_n b_n \sin (\omega_d nt)
\end{equation}
Writing the trigonometric functions  as complex exponentials, assuming $\omega '>0$ and if the SQUID plasma frequency is large ($\vert\omega\vert^2 C_J << 1$) we find

\begin{widetext}

\begin{equation}\label{pau}
\begin{split}
0&=a_\omega^{\text{in}}(1+ik_\omega L^0_{\text{eff}})+a_\omega^{\text{out}}(1-ik_\omega L^0_{\text{eff}})
+\sum_n \frac{a_n}{a_0} \bigg( \sqrt{\frac{\omega}{\omega- n\omega_d}}\theta(\omega- n\omega_d)(a_{\omega- n\omega_d}^{\text{in}}+a_{\omega- n\omega_d}^{\text{out}})\\
&+\sqrt{\frac{\omega}{n\omega_d -\omega}}\theta(n\omega_d -\omega)(a_{n\omega_d -\omega}^{\text{in}}+a_{n\omega_d -\omega}^{\text{out}})^\dagger +\sqrt{\frac{\omega}{\omega +n\omega_d}}(a_{\omega +n\omega_d}^{\text{in}}+a_{\omega +n\omega_d}^{\text{out}})\bigg) \\
&+\sum_n \frac{b_n}{a_0}\bigg(- \sqrt{\frac{\omega}{\omega -n\omega_d}}\theta(\omega -n\omega_d)(a_{\omega -n\omega_d}^{\text{in}}+a_{\omega -n\omega_d}^{\text{out}})-\sqrt{\frac{\omega}{n\omega_d -\omega}}\theta(n\omega_d -\omega)(a_{n\omega_d -\omega}^{\text{in}}+a_{n\omega_d -\omega}^{\text{out}})^\dagger\\
&+\sqrt{\frac{\omega}{\omega +n\omega_d}}(a_{\omega +n\omega_d}^{\text{in}}+a_{\omega +n\omega_d}^{\text{out}})\bigg) 
\end{split}
\end{equation}
\end{widetext}
where 
\begin{equation}\label{Leff0}
L_{\text{eff}}^0=\left(\frac{\Phi_0}{2\pi}\right)^2\frac{1}{L_0}\left(\frac{2}{a_0}\right)
\end{equation}
and we have set $\omega'\rightarrow\omega$.

Equation (\ref{pau}) is the general relation determining $a_\omega^{\text{out}}$ in terms of $a_\omega^{\text{in}}$ for an
arbitrary driving force.  We can  solve this equation perturbatively by writing 
\begin{equation}\label{perturbation}
a_\alpha^{\text{out}}=a_{\alpha 0}^{\text{out}}+\sum_n a_{\alpha n}^{\text{out}}\frac{a_n}{a_0}+\sum_n b_{\alpha n}^{\text{out}}\frac{b_n}{a_0 i}+O(2)
\end{equation}

where we require $\frac{a_n}{a_0}<< 1$ and $\frac{b_n}{b_0}<< 1$. Here $O(2)$ means second order in $a_n/a_0$ and $b_n/a_0$
With this, the 0th order term is
\begin{equation}\label{0order}
0=a_\omega^{\text{in}}(1+ik_\omega L^0_{\text{eff}})+a_{\omega 0}^{\text{out}}(1-ik_\omega L^0_{\text{eff}})
\end{equation}
yielding
\begin{equation}\label{a0out}
a_{\omega 0}^{\text{out}} = -\frac{1+ik_\omega L^0_{\text{eff}}}{1-ik_\omega L^0_{\text{eff}}} a_\omega^{\text{in}}  = R(\omega)a_\omega^{\text{in}}
\end{equation}

Using \eqref{a0out} and upon imposing the requirement that $k_\omega L^0_{\text{eff}}\ll 1$, which gives an upper bound on frequencies where our treatment is valid, the 1st order term is

\begin{widetext}
\begin{equation}\label{rosa}
\begin{split}
a_{\omega n}^{\text{out}}= \frac{2iL^0_{\text{eff}}}{v}& \left[\sqrt{\omega}\sqrt{\omega -n\omega_d}\theta(\omega -n\omega_d) e^{i(k_\omega+ k_{\omega -n\omega_d})L^0_{\text{eff}}}a_{\omega -n\omega_d}^{\text{in}}\right. \\
&\left. -\sqrt{\omega}\sqrt{n\omega_d -\omega}\theta(n\omega_d -\omega) e^{i(k_\omega- k_{n\omega_d -\omega})L^0_{\text{eff}}}a_{n\omega_d -\omega}^{in\dagger}\right.\\
&\left.+\sqrt{\omega}\sqrt{\omega +n\omega_d} e^{i(k_\omega+ k_{\omega +n\omega_d})L^0_{\text{eff}}}a_{\omega +n\omega_d}^{\text{in}}\right]
\end{split}
\end{equation}
and similarly
\begin{equation}\label{estrella}
\begin{split}
b_{\omega n}^{\text{out}}= \frac{2iL^0_{\text{eff}}}{v} &\left[-\sqrt{\omega}\sqrt{\omega -n\omega_d}\theta(\omega -n\omega_d) e^{i(k_\omega+ k_{\omega -n\omega_d})L^0_{\text{eff}}}a_{\omega -n\omega_d}^{\text{in}}\right.\\
&\left.+\sqrt{\omega}\sqrt{n\omega_d -\omega}\theta(n\omega_d -\omega) e^{i(k_\omega- k_{n\omega_d -\omega})L^0_{\text{eff}}}a_{n\omega_d -\omega}^{in\dagger}\right.\\
&\left.+\sqrt{\omega}\sqrt{\omega +n\omega_d} e^{i(k_\omega+ k_{\omega +n\omega_d})L^0_{\text{eff}}}a_{\omega +n\omega_d}^{\text{in}}\right]
\end{split}
\end{equation}
where we substituted $k_\alpha=\frac{\vert \alpha\vert}{v}$

Substituting Eqns. \eqref{a0out}, \eqref{rosa} and \eqref{estrella} into \eqref{perturbation}, we finally get
\begin{equation}\label{aout}
\begin{split}
a_\omega^{\text{out}}=R(\omega)a_\omega^{\text{in}}+\sum_n \bigg( &\left[\frac{a_n}{a_0}P(\omega ,\omega -n\omega_d)-i\frac{b_n}{a_0}P^*(\omega ,\omega -n\omega_d)\right]e^{i(k_\omega +k_{\omega -n\omega_d})L^0_{\text{eff}}} a_{\omega -n\omega_d}^{\text{in}}\\
+&\left[\frac{a_n}{a_0}P^*(\omega ,n \omega_d -\omega)-i\frac{b_n}{a_0}P(\omega ,n \omega_d -\omega)\right]e^{i(k_\omega -k_{n \omega_d -\omega})L^0_{\text{eff}}} a_{n \omega_d -\omega}^{in\dagger}\\
+&\left[\frac{a_n}{a_0}P(\omega ,\omega +n\omega_d)-i\frac{b_n}{a_0}P(\omega ,\omega +n\omega_d)\right]e^{i(k_\omega +k_{\omega +n\omega_d})L^0_{\text{eff}}} a_{\omega +n\omega_d}^{\text{in}}
\end{split}
\end{equation}
where we have defined
\begin{equation}\label{P}
P(\omega ',\omega '')=\frac{2iL^0_{\text{eff}}}{v}\sqrt{\omega '}\sqrt{\omega ''}\theta(\omega')\theta(\omega '')
\end{equation}
If the initial photon population of the field is given by that of a thermal bath of temperature $T$: $\bar{n}^{\text{in}}_\omega=\left(\exp(\hbar\omega/K_\beta T)-1\right)^{-1}$, then
\begin{equation}\label{nout3}
\begin{split}
\bar{n}^{\text{out}}_\omega =\vert R(\omega)\vert ^2 \bar{n}^{\text{in}}_\omega &+\frac{4 (L_{\text{eff}}^{0})^2}{v^2} \sum_n \bigg [ \omega(\vert\omega -n\omega_d\vert)\bigg \vert \frac{a_n}{a_0}+i\frac{b_n}{a_0}\bigg\vert^2 \bar{n}^{\text{in}}_{\vert\omega -n\omega_d\vert}\\
&+\omega(n\omega_d -\omega)\bigg\vert \frac{a_n}{a_0}+i\frac{b_n}{a_0}\bigg\vert^2 \Theta(n\omega_d -\omega)
+\omega(\omega +n\omega_d)\bigg\vert \frac{a_n}{a_0}-i\frac{b_n}{a_0}\bigg\vert^2 \bar{n}^{\text{in}}_{\omega + n\omega_d}\bigg ]
\end{split}
\end{equation}
Requiring that $k_B T<<\hbar \omega_d$, we neglect terms containing the small factor $\bar{n}^{\text{in}}_{\omega + n\omega_d}$, finally obtaining
\begin{equation}\label{noutfinal}
\bar{n}^{\text{out}}_\omega =\vert R(\omega)\vert ^2 \bar{n}^{\text{in}}_\omega +\frac{4 (L_{\text{eff}}^{0})^2}{v^2 \vert a_0 \vert^2} \sum_n \left[\vert a_n +i b_n\vert^2\left(\omega\vert\omega -n\omega_d\vert \bar{n}^{\text{in}}_{\vert\omega -n\omega_d\vert}+\omega(n\omega_d -\omega)\Theta(n\omega_d -\omega)\right)\right]
\end{equation}
\end{widetext}
and upon using 
\begin{eqnarray}
E_J(t) &=&\frac{a_0}{2}+\sum_n a_n \cos (\omega_d nt)+\sum_n b_n \sin (\omega_d nt)  \nonumber\\
&=& E_J^0+\delta E_J(t) \label{EJ}
\end{eqnarray}
we compute an effective length 
\begin{eqnarray}
L_{\text{eff}} &=&\left(\frac{\phi_0}{2\pi}\right)^2 \frac{1}{E_J(t)}\frac{1}{L_0}
 \nonumber\\
&=&\left(\frac{\phi_0}{2\pi}\right)^2 \frac{1}{E_J^0+\delta E_J(t)}\frac{1}{L_0}
 \nonumber\\
&\approx& L_{\text{eff}}^0-\delta L_{eff}\label{Leff}
\end{eqnarray}
with
\begin{equation}\label{deltaLeff}
\delta L_{\text{eff}}=L_{\text{eff}}^0\left(\frac{\delta E_J(t)}{E_J^0}\right)
\end{equation}

If we want to simulate a trajectory with a position given by $x=Z(t)$, then 
upon comparison with (\ref{abcmotion}) we obtain
\begin{equation}\label{xt}
z(t)= \delta L_{eff}  \Rightarrow \delta E_J(t)=\frac{E_J^0}{L_{\text{eff}}^0} z(t)
\end{equation}
and given $z(t)$ and its Fourier coefficients $\lbrace \tilde a_0,\tilde a_m, \tilde b_m\rbrace$  we find
\begin{equation}\label{FouCoeff}
\begin{split}
\tilde a_0=&0\\
\tilde a_m=&\frac{ 4}{a_0^2 L_0}\left(\frac{\phi_0}{2\pi} \right)^2 a_m\\
\tilde b_m=&\frac{ 4}{a_0^2 L_0}\left(\frac{\phi_0}{2\pi} \right)^2 b_m
\end{split}
\end{equation}

Recall that the external driving field as a function of the external flux is given by $E_J(t)=2E_J\vert\cos\left(\frac{\pi \phi_{\text{ext}}(t)}{\phi_0}\right)\vert$. Consequently
\begin{equation}\label{phit}
\phi_{\text{ext}}(t)=\frac{\phi_0}{\pi}\cos^{-1}\left(\frac{E_J(t)}{2E_J}\right)
\end{equation}
so the external flux as a function of the desired trajectory is 
\begin{equation}\label{external}
\phi_{\text{ext}}(t)=\frac{\phi_0}{\pi}\cos^{-1}\left(\frac{E_J^0}{2E_J}\left(1+\frac{ z(t)}{L_{\text{eff}}^0}\right)\right)
\end{equation}

\section{Parameters for Relativistic Trajectories} \label{NPRT} 

 In this section we  suggest physically relevant parameters  
for the relativistic trajectories SM, SA, and AUA (described in section \ref{RTMB}) and 
compute the number of photons produced for each.
From equations \eqref{Sinusoidal}, \eqref{Chen} and \eqref{AUA}, we see that each trajectory has a characteristic acceleration parameter (generically denoted $A$) that will roughly determine the scale of the proper acceleration, . Respectively $A$ is $R\omega_d^2$ for SM, $\alpha$ for SA and $a$ for AUA.

For each trajectory, the time averaged proper acceleration is  
\begin{equation}\label{averagea}
\bar{a}=\frac{\displaystyle{\int_{\tau(t=0)}^{\tau(t=2\pi/\omega_d)}\!\!\!\!\!\! \!\!\!\!\!\! 
\text{d}\tau\,a(\tau)}}{\displaystyle{\int_{\tau(t=0)}^{\tau\left(t=2\pi/\omega_d\right)}\!\!\!\!\!\! \!\!\!\!\!\! \!\!\!\!\!\! \text{d}\tau}}
\end{equation}
where $\tau$ is the proper time. We reproduce these in Table \ref{tabla1} for convenience. We notice that $\bar{a}$ is a monotonically increasing function of the accelerating parameter A for fixed frequency. For fixed A and varying frequency, $\bar{a}$ is monotonically increasing for SM and monotonically decreasing for SA. 
\begin{table}[h]
\begin{center}
  \begin{tabular}{| l | c | c | c |}
    \hline
     & SM & SA & AUA \\ \hline
    $\bar{a}$ & $\frac{v \omega_d \tanh^{-1} \frac{R\omega_d}{v}}{E\left(\frac{R^2\omega^2_d}{v^2}\right)}$ & $\frac{v\omega_d\sinh^{-1}\left(2\frac{\alpha}{v\omega_d}\right)}{F\left(\frac{\pi}{2},-4\left(\frac{\alpha}{v\omega_d}\right)^2\right)}$ & $a$ \\
    \hline
  \end{tabular} 
  \end{center}
\caption{Time averaged proper accelerations for the three trajectories studied.
$F(\phi,m)$ and $E(\phi,m)$ are  elliptic integrals of the first and second kind  respectively.}
\label{tabla1}
\end{table}

 As an estimator of how relativistic the trajectory is we can compare $\bar{a} t$ with the effective speed of light $v$. If $\bar{a}t\lesssim v$ the trajectory would be significantly relativistic. Since $v=\frac{2}{5}c$ this means that  $\bar{a}t=\frac{\bar{a}2\pi}{\omega_d} \approx \frac{2}{5}c$ or $\bar{a} \frac{2\pi}{\omega_d}\frac{5}{2 c} \approx 1$, so if $\frac{\omega_d}{2\pi}\sim O(10^{10})$ Hz, then $\bar{a}\sim O(10^{17})$ m/s$^2$. We remark that by  $\omega$, $\omega_d$ we mean angular frequencies, that is $2\pi \nu$, where $\nu$ is the linear frequency.

The values of the parameters employed in \cite{DCE2} are summarized in Table \ref{parametros}. These parameters yield a proper acceleration for the sinusoidal motion simulated in \cite{DCE2} of $\bar{a}=9.054\times10^{17}$ $\text{m}\,\text{s}^{-2}$, then $\bar{a} \frac{2\pi}{\omega_d}\frac{5}{2 c}=0.419$. For this acceleration and the oscillation period considered, neither the SA nor AUA  trajectories will yield any significant difference with the simple sinusoidal motion  as we will see   in section VI. 
\begin{table}[h!]
\begin{center}
  \begin{tabular}{| c | c |}
    \hline
     & SM \\ \hline
    $\frac{\omega_d}{2\pi}$ & $18\, \text{GHz}$  \\ \hline
    $E_J^0=\frac{a_0}{2}$ & $1.3 E_J$  \\ \hline
    $a_1$ & $\frac{\left(\frac{a_0}{2}\right)}{4}$  \\ \hline
    $E_J$ & $I_c \left(\frac{\phi_0}{2\pi}\right)$  \\ \hline
    $I_c$ & $1.25\,\mu \text{A}$  \\ \hline
    $C_J$ & $90\, \text{fF}$  \\ \hline
    $v$ & $.4 c$ \\ \hline
    $Z0$ & $55\, \Omega$ \\ \hline
    $\omega_s$ & $37.3\, \text{GHz}$ \\ \hline
  \end{tabular}
  \caption {Parameters used in \cite{DCE2}}
  \label{parametros}
\end{center}
\end{table}

In order to obtain  significant differences between the SM and the other two trajectories we need to work with larger $\bar{a}$ so as to reach speeds that are closer to the effective speed of light, and thus have larger contributions from higher than first order Fourier coefficients in \eqref{FouGen}.  We are constrained by the fact that the speed of the wall cannot be faster than the speed of light. Both the SA and AUA trajectories already incorporate this constraint by construction, but in the case of sinusoidal motion, not every value of the characteristic acceleration parameter is possible. In this case we will have the constraint
\begin{equation}\label{Rconstrain}
R\omega_d < {v}
\end{equation}
This means that for a maximum driving frequency of $\frac{\omega_d}{2\pi}=40$ GHz, and $v=.4 c$, then $R< .4775 \text{mm}$. 

We therefore impose three requirements in  choosing our  parameters. First, we set  $\bar{a}(A,\omega_d)=20\times10^{18}\ \text{m}\, \text{s}^{-2}$ and fix the same driving frequency $\omega_d$ for the trajectories. Next we select the characteristic acceleration parameter $A$ and driving frequency $\omega_d$  such that we retain the higher order contributions for the SA and AUA trajectories while ensuring that $\frac{E_J^0}{E_J} > 0.1$.   Finally, we  maximize the quantity $\frac{\bar{a}(A,\omega_d)}{\omega_d}$ so as to maximally amplify the contribution of the motions.
The first criterion  provides a point of comparison between trajectories, the second  gives a region on the plane ($A$,$\omega_d$) in which we can perform the experiment, and the third selects the parameters in which the trajectory is `maximally relativistic' given 
the other constraints.  

We find these criteria imply that  $\alpha=13.725\times 10^{18}$ $\text{m}\,\text{s}^{-2}$ for the SA motion and $a=20\times 10^{18}$ $\text{m}\,\text{s}^{-2}$ for the AUA motion and a driving frequency of $\frac{\omega_d}{2\pi}=14.6$ GHz for both trajectories. For the SM, in order to keep the driving frequency less than the plasma frequency and still achieve an average acceleration of $20\times10^{18}$ we would need $R \geq .398 \text{mm}$, and the greater the R, the smaller the required driving frequency. Due to Eq. \eqref{Rconstrain}, $R< .4775\text{mm}$, and to achieve the desired acceleration the minimum driving frequency is $\omega_d=31.7$ GHz. This driving frequency is much bigger than the frequency needed for SA and AUA. For this reason we shall first consider these two cases,  presenting the analogous results for the
sinusoidal case at the end of this work with the parameters used in \cite{DCE2} (presented in Table \ref{parametros}).  

We summarize in Table \ref{tabla2} the experimentally controlled parameters for the cases we subsequently analyze, unless otherwise specified.

\begin{table}[h!]
\begin{center}
  \begin{tabular}{| c | c | c |}
    \hline
     & SA & AUA \\ \hline
    $\bar{a}$ & $20\times 10^{18}\text{m}\,\text{s}^{-2}$ & $20\times 10^{18}\text{m}\,\text{s}^{-2}$ \\ \hline
    $A$ & $\alpha=13.725\times 10^{18}$ $\text{m}\,\text{s}^{-2}$ & $a=20\times 10^{18}$ $\text{m}\,\text{s}^{-2}$ \\ \hline
    $\omega_d/2\pi$ & $14.6\, \text{GHz}$ & $14.6\, \text{GHz}$ \\ \hline
    $E_J^0=\frac{a_0}{2}$ & $0.1002 E_J$ & $0.1006 E_J$ \\ \hline
    $a_1$ & $\frac{\left(\frac{a_0}{2}\right)}{4}$ & $\frac{\left(\frac{a_0}{2}\right)}{4}$ \\ \hline
    $E_J$ & $I_c \left(\frac{\phi_0}{2\pi}\right)$ & $I_c \left(\frac{\phi_0}{2\pi}\right)$ \\ \hline
    $I_c$ & $1.25\, \mu \text{A}$ & $1.25\, \mu \text{A}$  \\ \hline
    $C_J$ & $90\, \text{fF}$ & $90\, \text{fF}$ \\ \hline
    $v$ & $.4 c$ & $.4 c$ \\ \hline
    $Z0$ & $55\, \Omega$ & $55\, \Omega$ \\ \hline
    $\omega_s/2\pi$ & $37.3\, \text{GHz}$ & $37.3\, \text{GHz}$\\ \hline
  \end{tabular}
  \caption {Parameters used for each trajectory}
  \label{tabla2}
  \end{center}
\end{table}

\section{Results}\label{results}
 
\begin{figure}[h!]
\includegraphics[scale=.6]{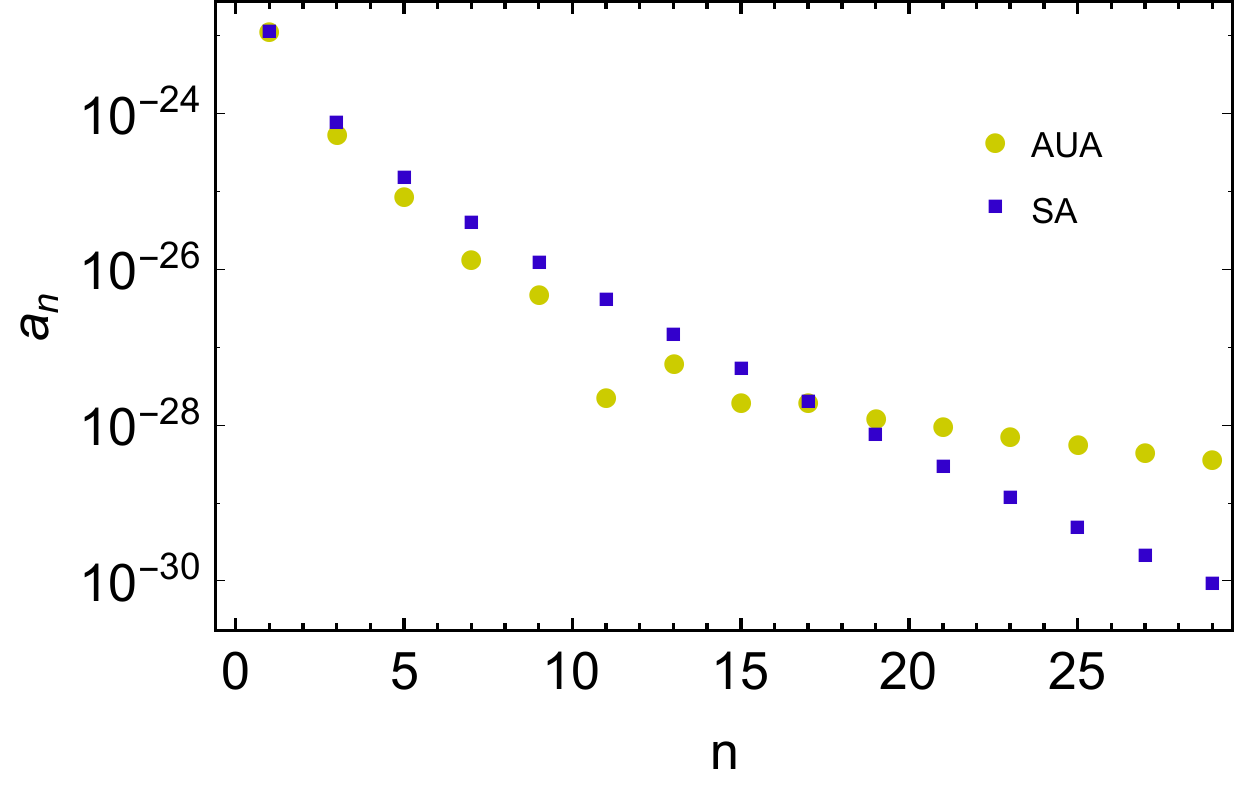} 
\caption{Fourier coefficients of the SA trajectory (squares, blue online) and AUA(circles, green online)} 
\label{fourier}
\label{an}
\end{figure}

With the parameters presented in Table \ref{tabla2}, the Fourier coefficients for the SA and AUA trajectories are non vanishing but are quickly suppressed as $n$ increases. We present them in Figure \ref{an}. In both cases we find that we get an exponential suppression, and so can safely consider only the first 3 Fourier coefficients.

Using equation \eqref{noutfinal} we can calculate the output number of photons as a function of the frequency $\omega$ and of the driving frequency $\omega_d$. Fixing the driving frequency as in Table \ref{tabla2}, we calculate $n_{\text{out}}(\omega)$ for different external temperatures. In Figure \ref{nout058}, we illustrate results for various values of the temperature of the thermal bath for each motion.  We see that  second order contributions are in principle detectable, as depicted in the insets.

\begin{figure*}
\includegraphics[scale=1.15]{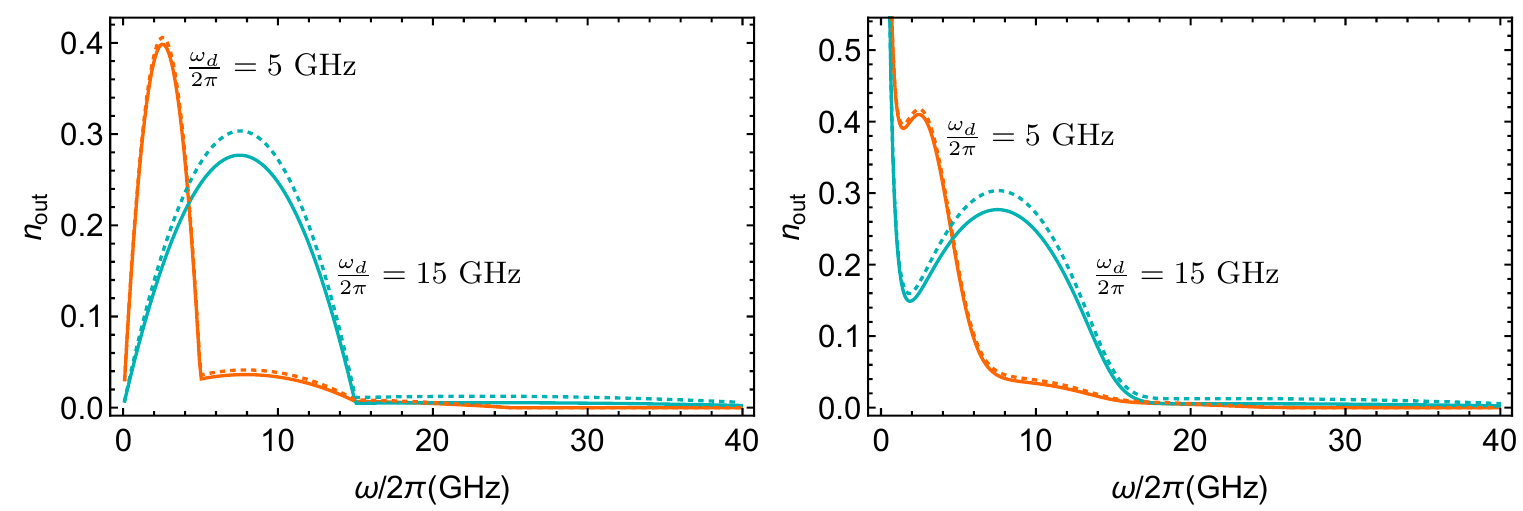}
\caption{Output number of photons for varying frequency $\omega$ and fixed driving frequencies $\omega_d$ as indicated in the figure. The solid lines correspond to AUA trajectory and the dashed lines to SA trajectory, where both have the same average acceleration $\bar{a}=20\times10^{18}\text{m}/\text{s}^{-2}$ for $\omega_d/2\pi=15\text{ GHz}$ and $\bar{a}=21.9\times10^{18}\text{m}/\text{s}^{-2}$ for $\omega_d/2\pi=5\text{ GHz}$, and for T=0 K (left) and T=25 mK (right), where $T$ is the temperature of the thermal bath.}
\label{noutw}
\end{figure*}

\begin{figure*}
\includegraphics[scale=1.15]{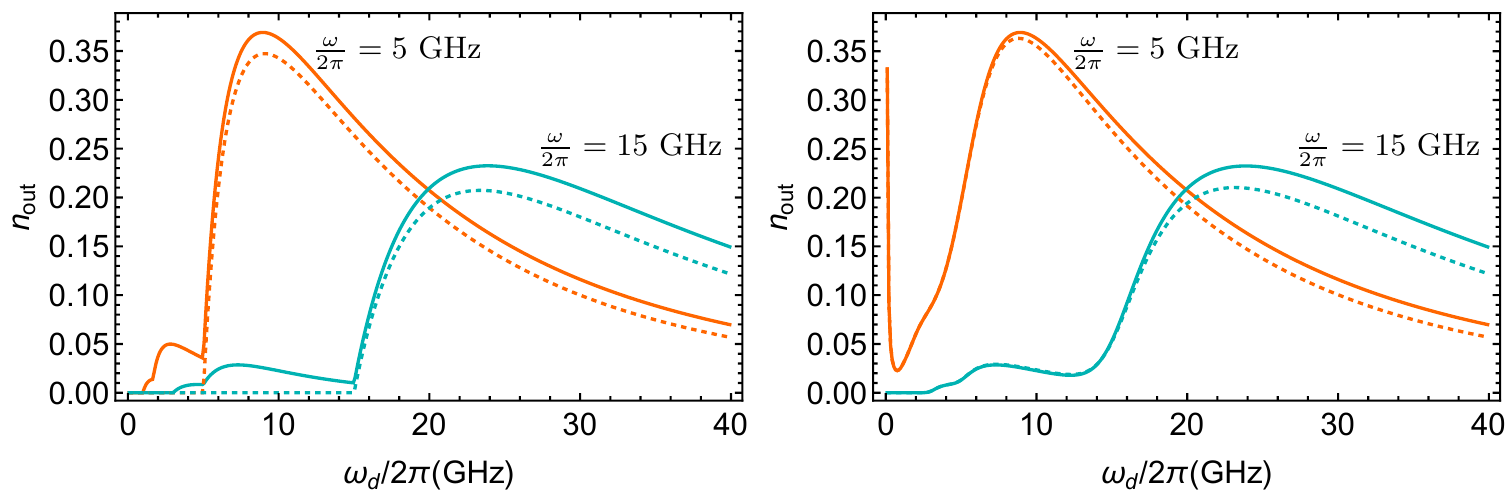}
\caption{Output number of photons for varying driving frequency $\omega_d$ and fixed frequencies 
$\omega$ as indicated in the figure. The solid lines correspond to AUA trajectory and the dashed lines to SA trajectory, where both have the same average acceleration $\bar{a} = 20 \times 10^{18}$ m/s$^2$ and for T=0 K (left) and T=25 mK (right), where $T$ is the temperature of the thermal bath.}
\label{noutwd}
\end{figure*}

In Figures  \ref{noutw} and \ref{noutwd} we calculate $n_{\text{out}}(\omega)$ (for two different fixed driving frequencies) and $n_{\text{out}}(\omega_d)$ (for two different fixed frequencies) respectively for two different temperatures of the thermal bath. For comparison purposes,  we present both trajectories together. We can see that even though small, there is a difference in the statistics for different trajectories. 

We see from Figures \ref{noutw} and \ref{noutwd} that the different relativistic motions are indeed distinguishable from their spectrum, with the distinction becoming more pronounced at larger values of $\omega_d$.  The maximum of the curve in figure \ref{noutw} occurs at values $\omega = n\omega_d/2 $.   An analytic expression for determining the maxima of the curves in figure \ref{noutwd} can be given in terms of elliptic functions; we shall not reproduce it here.

\begin{figure*}
\includegraphics[width=0.5\textwidth]{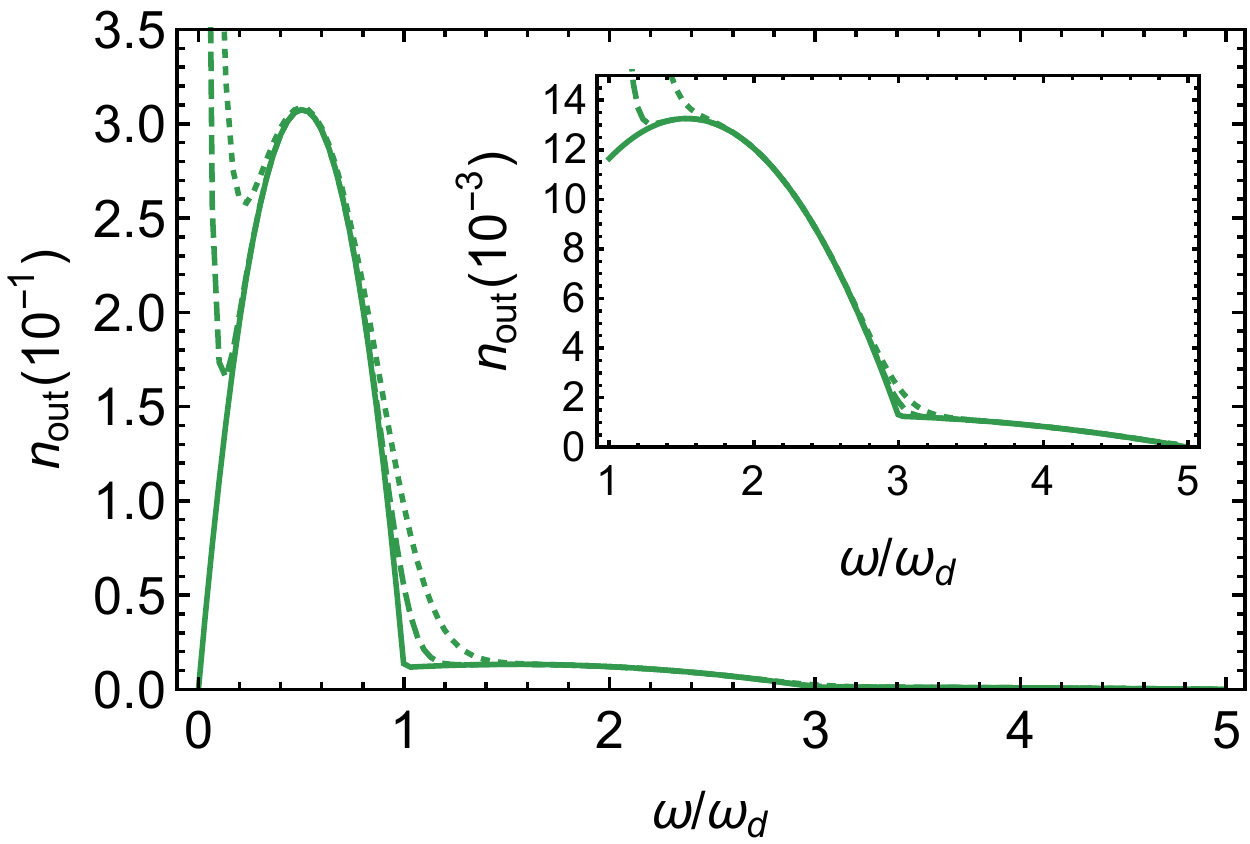}\includegraphics[width=0.5\textwidth]{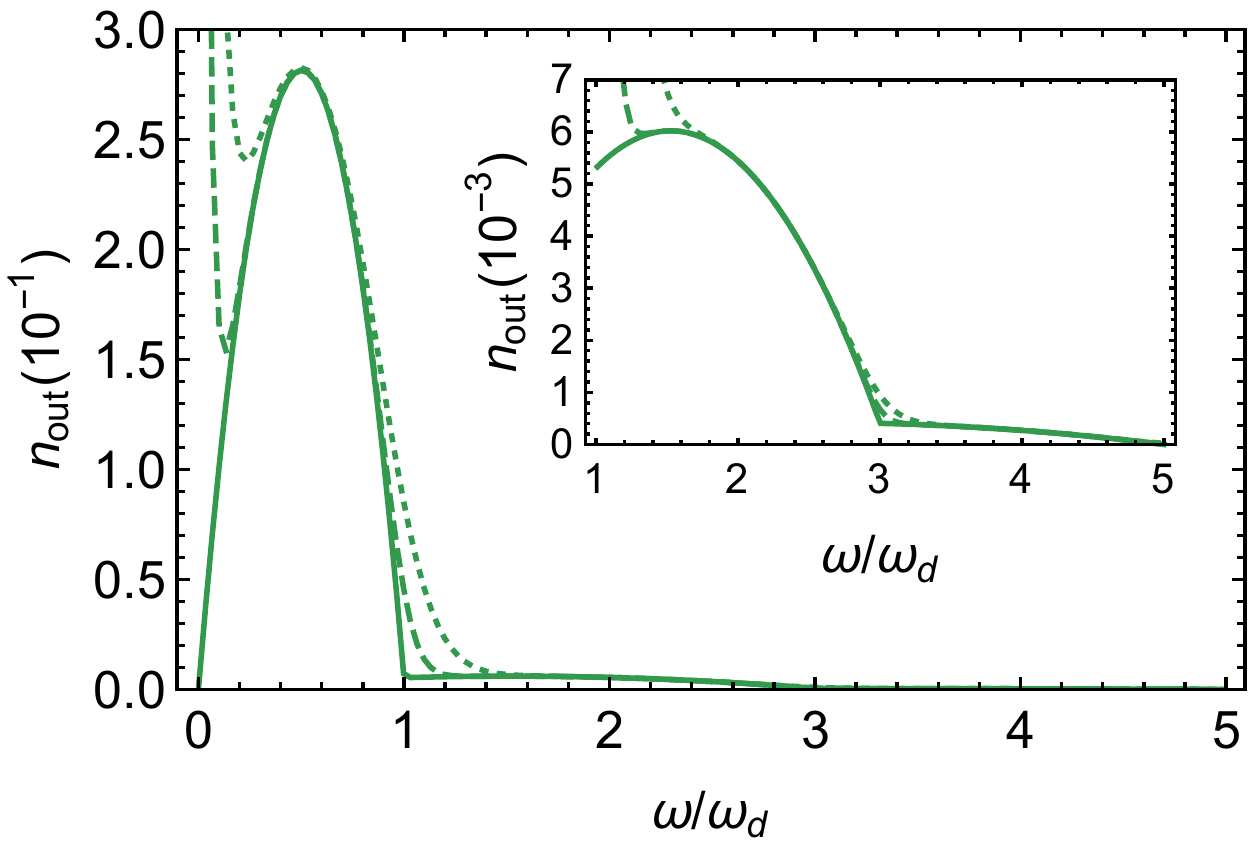}
\label{nout058}
\caption{Plots comparing $n_{\text{out}}$ at differing thermal bath temperatures $T=0$K (solid), $T=25$ mK(dashed) and $T=50$ mK(dotted) and fixed $\omega_d/2\pi=14.6\text{GHz}$ as a function of $\frac{\omega}{\omega_d}$ for SA trajectory (left) and AUA trajectory (right). The average acceleration for both trajectories is $\bar{a}=20\times 10^{18} \text{ m}/\text{s}^2$. The insets show detail for the second maximum.}
\label{nout058}
\end{figure*}

Finally we compute the output number of photons as a function of $\bar{a}$. Note that for AUA, the average acceleration is only a function of the acceleration parameter while for both SM and SA, due to the relativistic nature of the trajectories, the average proper acceleration depends nontrivially on the characteristic acceleration parameter and the driving frequency (periodicity of the motion). 

\begin{figure*}
\includegraphics[scale=1.15]{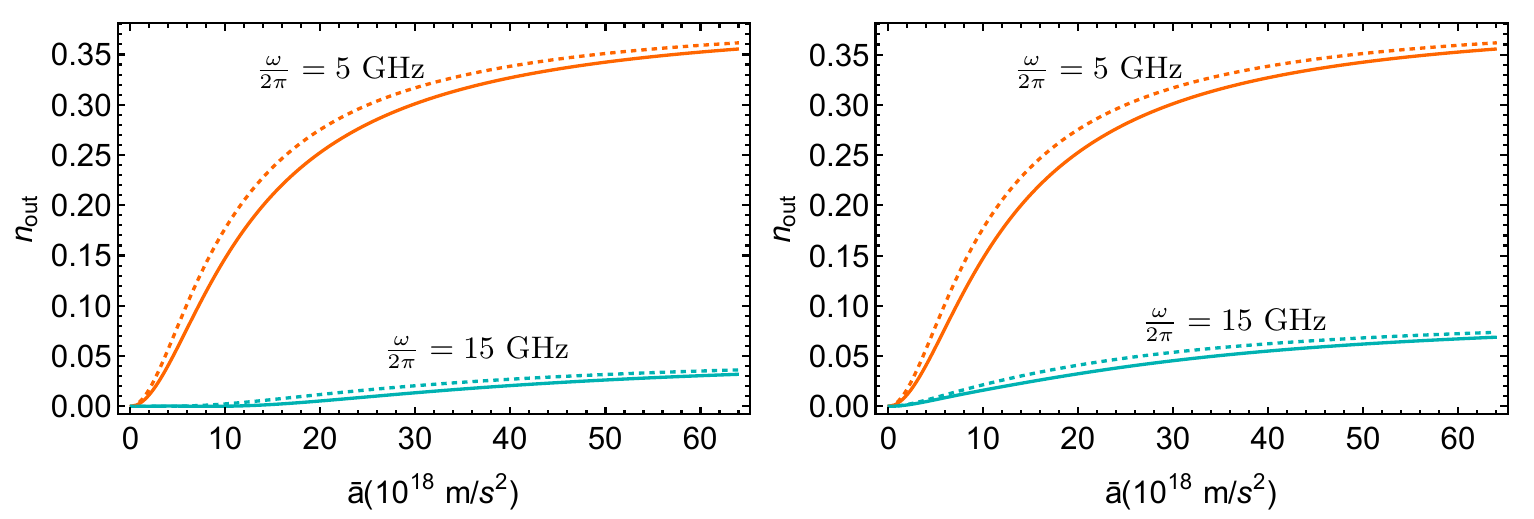}
\caption{Output number of photons for varying $\bar{a}$ for fixed driving frequency $\frac{\omega_d}{2\pi}=14.6 \text{GHz}$ and varying the characteristic acceleration parameter A for different frequencies $\omega$ as indicated in the plots, where the solid lines correspond to AUA trajectory and the dashed lines to SA trajectory,  at T=0 mK (left) and T=25 mK (right), where $T$ is the temperature of the thermal bath.}
\label{teffa}
\end{figure*}

Consequently there are two ways of having a variation in the acceleration. We can either  fix the characteristic acceleration parameter and vary the driving frequency $\omega_d$ or we can fix the driving frequency $\omega_d$  and vary the characteristic acceleration parameter. We will consider the case where we vary the acceleration by varying the acceleration parameter $A$, since this is the variable that carries the units of acceleration. The result is presented in Figure \ref{teffa}, where we set the driving frequency to $\frac{\omega_d}{2\pi}=14.6$ GHz. We notice that the output number of particles is an increasing monotonic function of the average acceleration.

\section{Sinusoidal Motion and the Dynamical Casimir Effect}

Turning now to the SM case, this is essentially the same as that considered in the dynamical Casimir effect  (DCE)\cite{Wilson,DCE,DCE2}.
To order $R\omega_d$ we are unable to produce any distinctly relativistic effects for this motion as per the discussion in section \ref{sect-cQED}.  As such, the DCE provides a cross-check on our approach.
 We set all the parameters to be the same as specified in \cite{DCE2} (presented in Table \ref{parametros}). These parameters give an effective length $L_{\text{eff}}^0=.44$ mm and a modulation $R=\delta L_{eff}=.11$ mm. With these parameters, we obtain the output number of photons calculated in \cite{DCE2}.
 
\begin{figure}[h!]
\includegraphics[scale=.5]{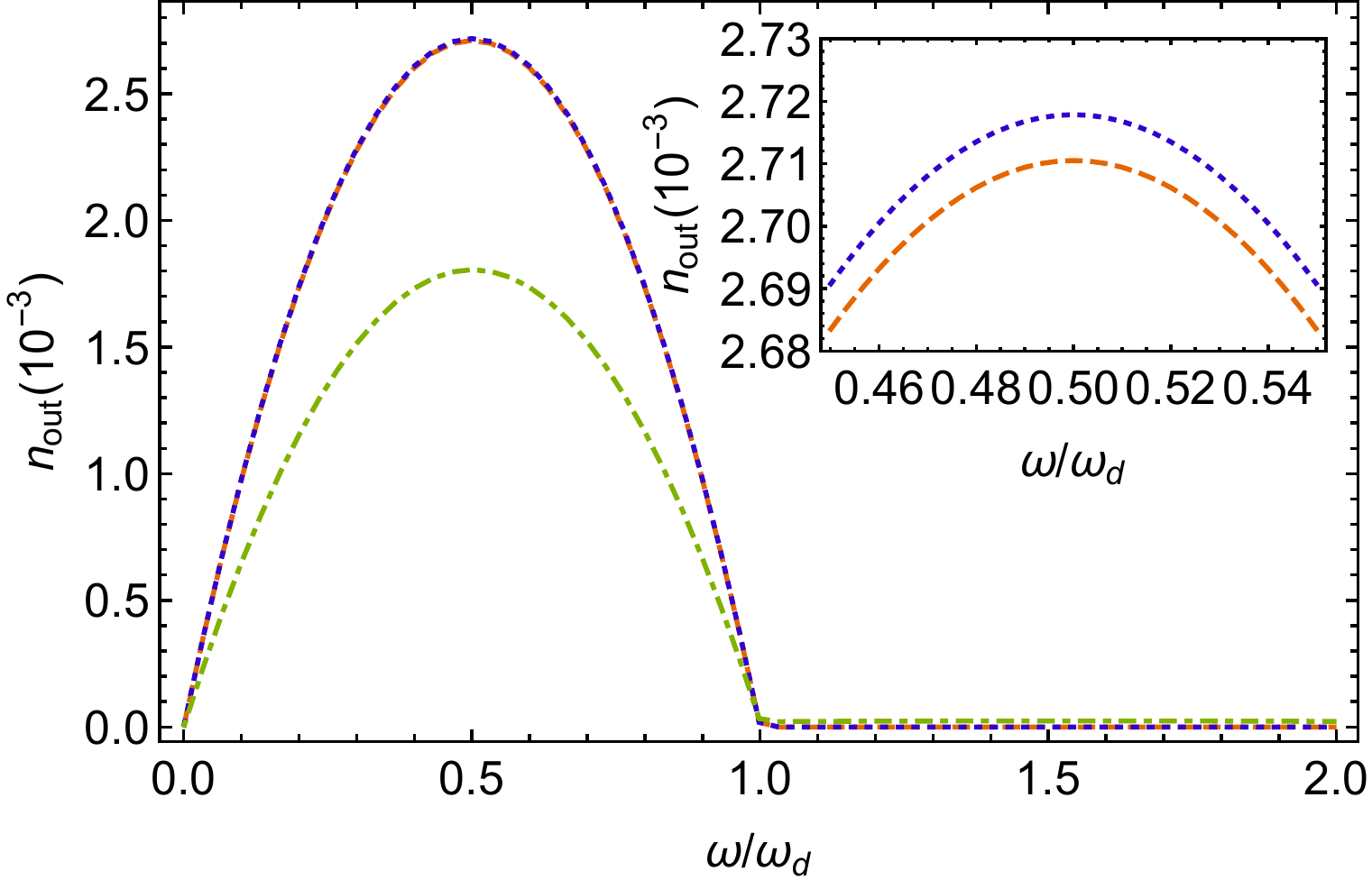} 
\caption{Plots comparing $n_{\text{out}}$ at a thermal bath temperature of $T=0$K and fixed $\omega_d/2\pi=18\text{GHz}$ as a function of $\frac{\omega}{\omega_d}$ for Sinusoidal trajectory (red-dashed), SA trajectory (blue-dotted) and AUA trajectory (green-dotdashed). The average acceleration for all the motions is $\bar{a}=9.054\times10^{17}$ $\text{m}\,\text{s}^{-2}$. The inset shows detail for the difference between Sinusoidal and SA.} 
\label{las3}
\end{figure}

\begin{figure}[h!]
\includegraphics[scale=.6]{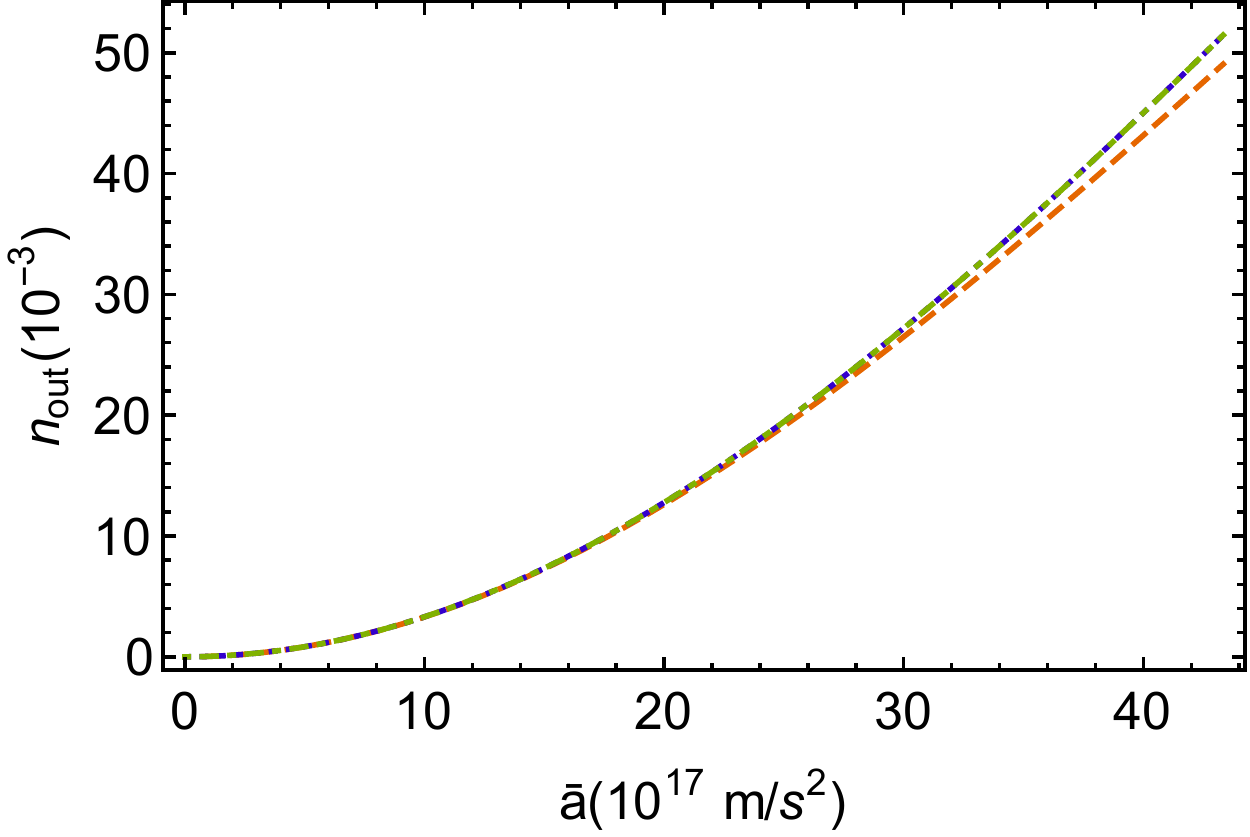} 
\caption{Plots comparing $n_{\text{out}}$ at a thermal bath temperature of $T=0$K and fixed frequency $\omega=9\text{GHz}$ as a function of $\bar{a}$ where we fix $\omega_d=18\text{GHz}$ for Sinusoidal trajectory (red-dashed), SA trajectory (blue-dotted) and AUA trajectory (green-dotdashed).} 
\label{las32}
\end{figure}

To compare the three trajectories, we set the driving frequency for SA and AUA to be the same as the Sinusoidal case and we modulate the acceleration parameter such that the average acceleration is the same for the three of them, which is $\bar{a}=9.054\times10^{17}$ $\text{m}\,\text{s}^{-2}$ as in \cite{DCE2}. In Figure \ref{las3} we present the result for $n_{\text{out}}(\omega)$ as a function of $\frac{\omega}{\omega_d}$ and in Figure \ref{las32} 
for $n_{\text{out}}$ as a function of $\bar{a}$. We notice that for this relatively small value of the acceleration, the output photon spectra for SA and SM is very similar, whereas the spectra for AUA is smaller. We also notice that the additional Fourier coefficients make noticeable changes in the output spectra only for higher values of the acceleration, as indicated in Figure \ref{nout058} and in contrast to Figure \ref{las3}.

\section{Conclusion}

  We have seen that the dynamical Casimir effect yields different particle creation distributions  depending on the trajectory of the moving  boundary condition.  
Despite the limitations concerning the Dirichlet boundary condition inherent to the cQED implementation that we point out in this manuscript, we have shown that a simulation of this effect in a superconducting circuit can distinguish different particle creation spectra due to different kinds of relativistic oscillatory motion (all of them yielding very similar periodic boundary trajectories as shown in Figure \ref{doukas}). We have shown that the simulation of these boundary trajectories is experimentally attainable with state of the art technology.

To relate our results to the phenomenology of the Unruh effect  we can associate the average number of observed particles created by the time-dependence of the boundaries to a temperature estimator. This can be done by relating the observed output flux density to $n_\text{out}$ in the same way as in \cite{DCE2}. Doing so yields a temperature estimator proportional to average number of created particles $T \propto \hbar \omega k_\textsc{b}n_{\text{out}}$. This temperature estimator  can be compared with the temperature perceived by an accelerated Unruh-DeWitt detector following the same trajectories we impose in our moving boundaries. 

These  results may be  helpful in shedding some light on a long debated question: How much can the dynamical Casimir effect be discussed in terms of the same physical phenomena behind the Unruh effect as seen by a freely accelerating particle detector?  One might argue that all moving boundary condition effects are basically manifestations of the DCE, and as such this should also be the case of an accelerated atom.  However the point of this study is the acceleration of the moving boundary conditions, and whether or not this picks up new features of the type expected from the Unruh effect for particle detectors with the same trajectories as studied in \cite{Doukas:2013noa}. As we can see from our results, the temperature estimator does not really follow the simple behaviour of the response of particle detectors predicted in \cite{Doukas:2013noa}, which may be suggesting that, beyond constant acceleration, the DCE may not be so easy to relate to the Unruh effect, possibly because of these nonequilibrium effects showing up in very different ways for particle detectors and accelerating mirrors. 

\section*{Acknowledgements}
E. M-M and R.B.M are supported by the NSERC Discovery Programme. P. C-U gratefully acknowledges funding from CONACYT. C.M.W acknowledges support from Industry Canada and the Government of Ontario. 

\newpage
\newpage
\bibliography{QCvRM}

\begin{thebibliography}{21}%
\makeatletter
\providecommand \@ifxundefined [1]{%
 \@ifx{#1\undefined}
}%
\providecommand \@ifnum [1]{%
 \ifnum #1\expandafter \@firstoftwo
 \else \expandafter \@secondoftwo
 \fi
}%
\providecommand \@ifx [1]{%
 \ifx #1\expandafter \@firstoftwo
 \else \expandafter \@secondoftwo
 \fi
}%
\providecommand \natexlab [1]{#1}%
\providecommand \enquote  [1]{``#1''}%
\providecommand \bibnamefont  [1]{#1}%
\providecommand \bibfnamefont [1]{#1}%
\providecommand \citenamefont [1]{#1}%
\providecommand \href@noop [0]{\@secondoftwo}%
\providecommand \href [0]{\begingroup \@sanitize@url \@href}%
\providecommand \@href[1]{\@@startlink{#1}\@@href}%
\providecommand \@@href[1]{\endgroup#1\@@endlink}%
\providecommand \@sanitize@url [0]{\catcode `\\12\catcode `\$12\catcode
  `\&12\catcode `\#12\catcode `\^12\catcode `\_12\catcode `\%12\relax}%
\providecommand \@@startlink[1]{}%
\providecommand \@@endlink[0]{}%
\providecommand \url  [0]{\begingroup\@sanitize@url \@url }%
\providecommand \@url [1]{\endgroup\@href {#1}{\urlprefix }}%
\providecommand \urlprefix  [0]{URL }%
\providecommand \Eprint [0]{\href }%
\providecommand \doibase [0]{http://dx.doi.org/}%
\providecommand \selectlanguage [0]{\@gobble}%
\providecommand \bibinfo  [0]{\@secondoftwo}%
\providecommand \bibfield  [0]{\@secondoftwo}%
\providecommand \translation [1]{[#1]}%
\providecommand \BibitemOpen [0]{}%
\providecommand \bibitemStop [0]{}%
\providecommand \bibitemNoStop [0]{.\EOS\space}%
\providecommand \EOS [0]{\spacefactor3000\relax}%
\providecommand \BibitemShut  [1]{\csname bibitem#1\endcsname}%
\let\auto@bib@innerbib\@empty
\bibitem [{\citenamefont {Birrell}\ and\ \citenamefont
  {Davies}(1984)}]{Birrell1984}%
  \BibitemOpen
  \bibfield  {author} {\bibinfo {author} {\bibfnamefont {N.}~\bibnamefont
  {Birrell}}\ and\ \bibinfo {author} {\bibfnamefont {P.}~\bibnamefont
  {Davies}},\ }\href@noop {} {\emph {\bibinfo {title} {Quantum fields in curved
  space}}}\ (\bibinfo  {publisher} {Cambridge university press},\ \bibinfo
  {year} {1984})\BibitemShut {NoStop}%
\bibitem [{\citenamefont {Unruh}(1976)}]{Unruh1976}%
  \BibitemOpen
  \bibfield  {author} {\bibinfo {author} {\bibfnamefont {W.}~\bibnamefont
  {Unruh}},\ }\href@noop {} {\bibfield  {journal} {\bibinfo  {journal}
  {Physical Review D}\ }\textbf {\bibinfo {volume} {14}},\ \bibinfo {pages}
  {870} (\bibinfo {year} {1976})}\BibitemShut {NoStop}%
\bibitem [{\citenamefont {Sewell}(1982)}]{Sewell1982}%
  \BibitemOpen
  \bibfield  {author} {\bibinfo {author} {\bibfnamefont {G.~L.}\ \bibnamefont
  {Sewell}},\ }\href {\doibase http://dx.doi.org/10.1016/0003-4916(82)90284-6}
  {\bibfield  {journal} {\bibinfo  {journal} {Annals of Physics}\ }\textbf
  {\bibinfo {volume} {141}},\ \bibinfo {pages} {201 } (\bibinfo {year}
  {1982})},\ \bibinfo {note} {department of Physics, Queen Mary College, Mile
  End Road, London \{E1\} 4NS, England}\BibitemShut {NoStop}%
\bibitem [{\citenamefont {Brown}\ \emph {et~al.}(2013)\citenamefont {Brown},
  \citenamefont {Mart\'in-Mart\'inez}, \citenamefont {Menicucci},\ and\
  \citenamefont {Mann}}]{Brown2012}%
  \BibitemOpen
  \bibfield  {author} {\bibinfo {author} {\bibfnamefont {E.~G.}\ \bibnamefont
  {Brown}}, \bibinfo {author} {\bibfnamefont {E.}~\bibnamefont
  {Mart\'in-Mart\'inez}}, \bibinfo {author} {\bibfnamefont {N.~C.}\
  \bibnamefont {Menicucci}}, \ and\ \bibinfo {author} {\bibfnamefont {R.~B.}\
  \bibnamefont {Mann}},\ }\href {\doibase 10.1103/PhysRevD.87.084062}
  {\bibfield  {journal} {\bibinfo  {journal} {Phys. Rev. D}\ }\textbf {\bibinfo
  {volume} {87}},\ \bibinfo {pages} {084062} (\bibinfo {year} {2013})},\
  \Eprint {http://arxiv.org/abs/1212.1973} {1212.1973} \BibitemShut {NoStop}%
\bibitem [{\citenamefont {Bruschi}\ \emph {et~al.}(2012)\citenamefont
  {Bruschi}, \citenamefont {Lee},\ and\ \citenamefont {Fuentes}}]{Bruschi2012}%
  \BibitemOpen
  \bibfield  {author} {\bibinfo {author} {\bibfnamefont {D.~E.}\ \bibnamefont
  {Bruschi}}, \bibinfo {author} {\bibfnamefont {A.~R.}\ \bibnamefont {Lee}}, \
  and\ \bibinfo {author} {\bibfnamefont {I.}~\bibnamefont {Fuentes}},\
  }\href@noop {} {\  (\bibinfo {year} {2012})},\ \Eprint
  {http://arxiv.org/abs/1212.2110} {1212.2110} \BibitemShut {NoStop}%
\bibitem [{\citenamefont {Hu}\ \emph {et~al.}(2012)\citenamefont {Hu},
  \citenamefont {Lin},\ and\ \citenamefont {Louko}}]{Hu:2012jr}%
  \BibitemOpen
  \bibfield  {author} {\bibinfo {author} {\bibfnamefont {B.}~\bibnamefont
  {Hu}}, \bibinfo {author} {\bibfnamefont {S.-Y.}\ \bibnamefont {Lin}}, \ and\
  \bibinfo {author} {\bibfnamefont {J.}~\bibnamefont {Louko}},\ }\href
  {\doibase 10.1088/0264-9381/29/22/224005} {\bibfield  {journal} {\bibinfo
  {journal} {Class.Quant.Grav.}\ }\textbf {\bibinfo {volume} {29}},\ \bibinfo
  {pages} {224005} (\bibinfo {year} {2012})},\ \Eprint
  {http://arxiv.org/abs/1205.1328} {arXiv:1205.1328 [quant-ph]} \BibitemShut
  {NoStop}%
\bibitem [{\citenamefont {Doukas}\ \emph {et~al.}(2013)\citenamefont {Doukas},
  \citenamefont {Lin}, \citenamefont {Hu},\ and\ \citenamefont
  {Mann}}]{Doukas:2013noa}%
  \BibitemOpen
  \bibfield  {author} {\bibinfo {author} {\bibfnamefont {J.}~\bibnamefont
  {Doukas}}, \bibinfo {author} {\bibfnamefont {S.-Y.}\ \bibnamefont {Lin}},
  \bibinfo {author} {\bibfnamefont {B.}~\bibnamefont {Hu}}, \ and\ \bibinfo
  {author} {\bibfnamefont {R.~B.}\ \bibnamefont {Mann}},\ }\href {\doibase
  10.1007/JHEP11(2013)119} {\bibfield  {journal} {\bibinfo  {journal} {JHEP}\
  }\textbf {\bibinfo {volume} {1311}},\ \bibinfo {pages} {119} (\bibinfo {year}
  {2013})},\ \Eprint {http://arxiv.org/abs/1307.4360} {arXiv:1307.4360}
  \BibitemShut {NoStop}%
\bibitem [{\citenamefont {Mann}\ and\ \citenamefont
  {Villalba}(2009)}]{Mann:2009dma}%
  \BibitemOpen
  \bibfield  {author} {\bibinfo {author} {\bibfnamefont {R.~B.}\ \bibnamefont
  {Mann}}\ and\ \bibinfo {author} {\bibfnamefont {V.~M.}\ \bibnamefont
  {Villalba}},\ }\href {\doibase 10.1103/PhysRevA.80.022305} {\bibfield
  {journal} {\bibinfo  {journal} {Phys.Rev.}\ }\textbf {\bibinfo {volume}
  {A80}},\ \bibinfo {pages} {022305} (\bibinfo {year} {2009})},\ \Eprint
  {http://arxiv.org/abs/0902.1580} {arXiv:0902.1580 [quant-ph]} \BibitemShut
  {NoStop}%
\bibitem [{\citenamefont {Ostapchuk}\ \emph {et~al.}(2012)\citenamefont
  {Ostapchuk}, \citenamefont {Lin}, \citenamefont {Mann},\ and\ \citenamefont
  {Hu}}]{Ostapchuk:2011ud}%
  \BibitemOpen
  \bibfield  {author} {\bibinfo {author} {\bibfnamefont {D.~C.}\ \bibnamefont
  {Ostapchuk}}, \bibinfo {author} {\bibfnamefont {S.-Y.}\ \bibnamefont {Lin}},
  \bibinfo {author} {\bibfnamefont {R.~B.}\ \bibnamefont {Mann}}, \ and\
  \bibinfo {author} {\bibfnamefont {B.}~\bibnamefont {Hu}},\ }\href {\doibase
  10.1007/JHEP07(2012)072} {\bibfield  {journal} {\bibinfo  {journal} {JHEP}\
  }\textbf {\bibinfo {volume} {1207}},\ \bibinfo {pages} {072} (\bibinfo {year}
  {2012})},\ \Eprint {http://arxiv.org/abs/1108.3377} {arXiv:1108.3377 [gr-qc]}
  \BibitemShut {NoStop}%
\bibitem [{\citenamefont {Wilson}\ \emph {et~al.}(2011)\citenamefont {Wilson},
  \citenamefont {Johansson}, \citenamefont {Pourkabirian}, \citenamefont
  {Simoen}, \citenamefont {Johansson}, \citenamefont {Duty}, \citenamefont
  {Nori},\ and\ \citenamefont {Delsing}}]{Wilson}%
  \BibitemOpen
  \bibfield  {author} {\bibinfo {author} {\bibfnamefont {C.~M.}\ \bibnamefont
  {Wilson}}, \bibinfo {author} {\bibfnamefont {G.}~\bibnamefont {Johansson}},
  \bibinfo {author} {\bibfnamefont {A.}~\bibnamefont {Pourkabirian}}, \bibinfo
  {author} {\bibfnamefont {M.}~\bibnamefont {Simoen}}, \bibinfo {author}
  {\bibfnamefont {J.~R.}\ \bibnamefont {Johansson}}, \bibinfo {author}
  {\bibfnamefont {T.}~\bibnamefont {Duty}}, \bibinfo {author} {\bibfnamefont
  {F.}~\bibnamefont {Nori}}, \ and\ \bibinfo {author} {\bibfnamefont
  {P.}~\bibnamefont {Delsing}},\ }\href@noop {} {\bibfield  {journal} {\bibinfo
   {journal} {Nature}\ }\textbf {\bibinfo {volume} {479}},\ \bibinfo {pages}
  {376} (\bibinfo {year} {2011})}\BibitemShut {NoStop}%
\bibitem [{\citenamefont {Brenna}\ \emph {et~al.}(2013)\citenamefont {Brenna},
  \citenamefont {Brown}, \citenamefont {Mann},\ and\ \citenamefont
  {Mart\'in-Mart\'inez}}]{Brenna2013}%
  \BibitemOpen
  \bibfield  {author} {\bibinfo {author} {\bibfnamefont {W.~G.}\ \bibnamefont
  {Brenna}}, \bibinfo {author} {\bibfnamefont {E.~G.}\ \bibnamefont {Brown}},
  \bibinfo {author} {\bibfnamefont {R.~B.}\ \bibnamefont {Mann}}, \ and\
  \bibinfo {author} {\bibfnamefont {E.}~\bibnamefont {Mart\'in-Mart\'inez}},\
  }\href {\doibase 10.1103/PhysRevD.88.064031} {\bibfield  {journal} {\bibinfo
  {journal} {Phys. Rev. D}\ }\textbf {\bibinfo {volume} {88}},\ \bibinfo
  {pages} {064031} (\bibinfo {year} {2013})},\ \Eprint
  {http://arxiv.org/abs/1307.3335} {1307.3335} \BibitemShut {NoStop}%
\bibitem [{\citenamefont {Brenna}\ \emph {et~al.}(2015)\citenamefont {Brenna},
  \citenamefont {Mann},\ and\ \citenamefont
  {Martin-Martinez}}]{Brenna:2015fga}%
  \BibitemOpen
  \bibfield  {author} {\bibinfo {author} {\bibfnamefont {W.~G.}\ \bibnamefont
  {Brenna}}, \bibinfo {author} {\bibfnamefont {R.~B.}\ \bibnamefont {Mann}}, \
  and\ \bibinfo {author} {\bibfnamefont {E.}~\bibnamefont {Martin-Martinez}},\
  }\href@noop {} {\  (\bibinfo {year} {2015})},\ \Eprint
  {http://arxiv.org/abs/1504.02468} {arXiv:1504.02468 [quant-ph]} \BibitemShut
  {NoStop}%
\bibitem [{\citenamefont {Johansson}\ \emph {et~al.}(2009)\citenamefont
  {Johansson}, \citenamefont {Johansson}, \citenamefont {Wilson},\ and\
  \citenamefont {Nori}}]{DCE}%
  \BibitemOpen
  \bibfield  {author} {\bibinfo {author} {\bibfnamefont {J.~R.}\ \bibnamefont
  {Johansson}}, \bibinfo {author} {\bibfnamefont {G.}~\bibnamefont
  {Johansson}}, \bibinfo {author} {\bibfnamefont {C.~M.}\ \bibnamefont
  {Wilson}}, \ and\ \bibinfo {author} {\bibfnamefont {F.}~\bibnamefont
  {Nori}},\ }\href {\doibase 10.1103/PhysRevLett.103.147003} {\bibfield
  {journal} {\bibinfo  {journal} {Phys. Rev. Lett.}\ }\textbf {\bibinfo
  {volume} {103}},\ \bibinfo {pages} {147003} (\bibinfo {year}
  {2009})}\BibitemShut {NoStop}%
\bibitem [{\citenamefont {Johansson}\ \emph {et~al.}(2010)\citenamefont
  {Johansson}, \citenamefont {Johansson}, \citenamefont {Wilson},\ and\
  \citenamefont {Nori}}]{DCE2}%
  \BibitemOpen
  \bibfield  {author} {\bibinfo {author} {\bibfnamefont {J.~R.}\ \bibnamefont
  {Johansson}}, \bibinfo {author} {\bibfnamefont {G.}~\bibnamefont
  {Johansson}}, \bibinfo {author} {\bibfnamefont {C.~M.}\ \bibnamefont
  {Wilson}}, \ and\ \bibinfo {author} {\bibfnamefont {F.}~\bibnamefont
  {Nori}},\ }\href {\doibase 10.1103/PhysRevA.82.052509} {\bibfield  {journal}
  {\bibinfo  {journal} {Phys. Rev. A}\ }\textbf {\bibinfo {volume} {82}},\
  \bibinfo {pages} {052509} (\bibinfo {year} {2010})}\BibitemShut {NoStop}%
\bibitem [{\citenamefont {Chen}\ and\ \citenamefont
  {Tajima}(1999)}]{Chen1999a}%
  \BibitemOpen
  \bibfield  {author} {\bibinfo {author} {\bibfnamefont {P.}~\bibnamefont
  {Chen}}\ and\ \bibinfo {author} {\bibfnamefont {T.}~\bibnamefont {Tajima}},\
  }\href {\doibase 10.1103/PhysRevLett.83.256} {\bibfield  {journal} {\bibinfo
  {journal} {Phys. Rev. Lett.}\ }\textbf {\bibinfo {volume} {83}},\ \bibinfo
  {pages} {256} (\bibinfo {year} {1999})}\BibitemShut {NoStop}%
\bibitem [{\citenamefont {Moore}(1970)}]{Moore1}%
  \BibitemOpen
  \bibfield  {author} {\bibinfo {author} {\bibfnamefont {G.~T.}\ \bibnamefont
  {Moore}},\ }\href@noop {} {\bibfield  {journal} {\bibinfo  {journal} {J.
  Math. Phys.}\ }\textbf {\bibinfo {volume} {11}},\ \bibinfo {pages} {2679}
  (\bibinfo {year} {1970})}\BibitemShut {NoStop}%
\bibitem [{\citenamefont {Fulling}\ and\ \citenamefont
  {Davies}(1976)}]{Fulling1}%
  \BibitemOpen
  \bibfield  {author} {\bibinfo {author} {\bibfnamefont {S.~A.}\ \bibnamefont
  {Fulling}}\ and\ \bibinfo {author} {\bibfnamefont {P.~C.~W.}\ \bibnamefont
  {Davies}},\ }\href@noop {} {\bibfield  {journal} {\bibinfo  {journal} {Proc.
  R. Soc. London, Ser. A}\ }\textbf {\bibinfo {volume} {348}},\ \bibinfo
  {pages} {393} (\bibinfo {year} {1976})}\BibitemShut {NoStop}%
\bibitem [{Dod(2001)}]{Dodonov1}%
  \BibitemOpen
  \href@noop {} {\bibfield  {journal} {\bibinfo  {journal} {Adv. Chem. Phys.}\
  }\textbf {\bibinfo {volume} {119}},\ \bibinfo {pages} {309} (\bibinfo {year}
  {2001})}\BibitemShut {NoStop}%
\bibitem [{\citenamefont {Corona-Ugalde}\ \emph {et~al.}()\citenamefont
  {Corona-Ugalde}, \citenamefont {Louko},\ and\ \citenamefont {Mann}}]{inprep}%
  \BibitemOpen
  \bibfield  {author} {\bibinfo {author} {\bibfnamefont {P.}~\bibnamefont
  {Corona-Ugalde}}, \bibinfo {author} {\bibfnamefont {J.}~\bibnamefont
  {Louko}}, \ and\ \bibinfo {author} {\bibfnamefont {R.~B.}\ \bibnamefont
  {Mann}},\ }\href@noop {} {\enquote {\bibinfo {title} {In preparation},}\
  }\BibitemShut {NoStop}%
\bibitem [{\citenamefont {Dodonov}(2010)}]{generalizedDynamical}%
  \BibitemOpen
  \bibfield  {author} {\bibinfo {author} {\bibfnamefont {V.~V.}\ \bibnamefont
  {Dodonov}},\ }\href {http://stacks.iop.org/1402-4896/82/i=3/a=038105}
  {\bibfield  {journal} {\bibinfo  {journal} {Physica Scripta}\ }\textbf
  {\bibinfo {volume} {82}},\ \bibinfo {pages} {038105} (\bibinfo {year}
  {2010})}\BibitemShut {NoStop}%
\bibitem [{\citenamefont {Doukas}\ and\ \citenamefont {Louko}(2015)}]{JormJas}%
  \BibitemOpen
  \bibfield  {author} {\bibinfo {author} {\bibfnamefont {J.}~\bibnamefont
  {Doukas}}\ and\ \bibinfo {author} {\bibfnamefont {J.}~\bibnamefont {Louko}},\
  }\href {\doibase 10.1103/PhysRevD.91.044010} {\bibfield  {journal} {\bibinfo
  {journal} {Phys. Rev. D}\ }\textbf {\bibinfo {volume} {91}},\ \bibinfo
  {pages} {044010} (\bibinfo {year} {2015})}\BibitemShut {NoStop}%
\end{thebibliography}%
\end{document}